# Reflectionless Plasma Ignition via High-Power Virtual Perfect Absorption

Théo Delage[1,*], Jérôme Sokoloff[1], Olivier Pascal[1], Valentin Mazières[2], Alex Krasnok[3,*], and Thierry Callegari[1]

[1]LAPLACE, Université de Toulouse, CNRS, INPT, UPS, Toulouse, France

[2]ISAE-SUPAERO, Université de Toulouse, Toulouse, France

[3]Department of Electrical and Computer Engineering, Florida International University, Miami, FL 33174, USA

*To whom correspondence should be addressed: theo.delage@laplace.univ-tlse.fr, akrasnok@fiu.edu

**Abstract**

Plasma ignition is critical in various scientific and industrial applications, demanding an efficient and robust execution mechanism. In this work, we present an innovative approach to plasma ignition by incorporating the analysis of fundamental aspects of light scattering in the complex frequency plane. For the first time, we demonstrate the high-power virtual perfect absorption (VPA) regime, a groundbreaking method for perfectly capturing light within a resonator. By carefully designing the temporal profile of the incident wave, we effectively minimize reflections during the ignition stages, thereby significantly enhancing the efficiency and resilience of the process. Through comprehensive experimental investigations, we validate the viability of this approach, establishing VPA as a powerful tool for reflectionless excitation and optimal control of plasma discharge. By addressing the limitations of conventional plasma ignition methods, this research represents a pivotal step towards transformative advancements in plasma technology, with promising implications for improving the performance and sustainability of numerous applications.

## Introduction

Light absorption, a fundamental process associated with light-matter interactions, is a cornerstone for a broad spectrum of optical functionalities. These include a myriad of applications such as optical sensing, optical communication, biomedical imaging, photovoltaics, and energy transfer, each relying heavily on precisely controlling how light interacts with different materials. In this context, coherent perfect absorption (CPA) has significantly improved our control over this process by exciting a scattering matrix zero on the real frequency axis[1–3]. This approach enables the efficient development of optical switches and logic gates[4,5]. However, recent developments have also shown the potential of controlling optical phenomena in lossless systems, emulating a CPA-like response by manipulating complex frequency plane scattering matrix zeros giving rise to the concept of virtual perfect absorption (VPA)[6]. VPA allows temporary light storage in the system, which can be released on demand. This mechanism maintains a critical coupling in high-Q resonators, requiring a balance between directly reflected waves and leakage waves[6,7]. This balance is achieved using a sinusoidal incident wave with an exponentially increasing amplitude. The "virtual" in VPA indicates that it stores the incident wave's electromagnetic energy, contrasting with conventional absorbers, which convert this energy to other forms[1,2]. VPA has been experimentally demonstrated in various systems, including elastodynamic waves[8], water waves[9], lumped elements[10], and microwave resonant cavities[11]. It holds great potential for efficient light storage and release in optical and quantum memories. The discovery of this phenomenon has prompted a reevaluation of the core principles underlying light scattering[12]. As a result, researchers have recently uncovered several novel effects that were previously hindered in the complex frequency plane, including virtual critical coupling[7,11], unusual optical pulling force[13], virtual parity-time symmetry[14], and the non-Hermitian skin effect[15].

Nevertheless, the storage capacity of the electromagnetic field is bound by the physical limitations of the material filling the resonator. Once this material reaches its threshold, the field storage can no longer be sustained. This poses an intriguing question: *Could the VPA effect enable efficient and reflectionless plasma ignition within a gas-filled resonator?* Plasma, representing the ionized state of matter, is a crucial component in many industrial processes and cosmic phenomena[16–20]. In microelectronics, for instance, plasma plays a pivotal role in the etching process used for intricate circuit designs[21]. Moreover, it is indispensable in medicine, specifically in sterilization techniques used against pathogens, including the SARS-CoV-2 virus[22–24]. It also finds applications in propulsion technologies[25,26] and fusion power[27,28]. Therefore, finding an answer to this question could potentially trigger substantial progress in comprehending these applications and establish the foundation for creating new practical applications.

Among the numerous methods for efficient plasma ignition, electromagnetic energy stands out for its exceptional proficiency[29,30]. This approach concentrates electromagnetic waves intensely to trigger conditions conducive to plasma breakdown. Various components, including surfatrons[31], slot antennas[32], and resonant cavities, are employed in these systems, culminating in the development of resonator-based plasma sources (RPS)[33,34]. RPS systems bring unique advantages, such as the absence of electrodes, effectively preventing contamination, and remarkable energy efficiency across a wide range of gas pressures[35]. Of particular note is the performance of RPS systems at atmospheric pressure, where they exhibit prolonged lifetimes and stable discharge operations under diverse conditions[36]. This resilience under variable conditions highlights the potential of RPS systems, reinforcing their role and promise in advancing plasma technology.

Despite microwave sources' high efficiency for plasma ignition, enhancing these systems presents significant challenges. The highly dynamic electromagnetic conditions before and after plasma formation[37,38] often induce energy reflections that cause substantial energy loss and potential component damage in resonator-based plasma sources (RPS). Although isolators or circulators have been proposed to mitigate this problem, they prove inadequate due to their limited capacity to handle intense microwave fields. Current solutions propose using external electronic or mechanical dynamic circuits to minimize reflections during plasma formation and deactivation stages[20]. However, these strategies primarily focus on system adjustment rather than directly addressing the excitation method. We argue that considering the excitation approach could pave the way for a more efficient and resilient solution, which is crucial given the potential high-impact plasma technology applications across various fields.

In this work, we reveal a novel approach to plasma ignition by resonator excitation through a complex frequency signal. We specifically explore the high-power VPA regime, introducing, for the first time, an innovative method for close-to-perfect capturing high-power electromagnetic field within a resonator while intricately devising the temporal profile of the incident wave to minimize reflections during the ignition stages. Our results underscore the feasibility and efficiency of this approach, thus establishing VPA as a potent tool for reflectionless excitation and optimal plasma discharge control. By identifying and addressing the limitations of traditional plasma ignition methods, our work sets a substantial foundation for enhancing plasma ignition systems, marking a significant stride towards transformative advancements in plasma technology with promising sustainability implications.

**Results and discussion**

Plasma ignition demands a high microwave electric field, typically around $10^4$ V/m for argon gas[39]. An electromagnetic cavity can facilitate such high field levels in microwaves. At resonance frequencies, the waves reflected within the cavity interfere constructively, forming an eigenmode, thereby accumulating electromagnetic energy[40,41]. This energy influx can occur through multiple or single access ports. In the simplest scenario with a single access port, the energy stored in the cavity is the incoming power at the access minus the outgoing power and losses at the cavity walls. These losses depend on the material's conductivity and the specific cavity mode. Outgoing power is influenced by the direct reflection of incident waves at the access port and the escape of waves contained in the cavity[40,41]. By adjusting the cavity's coupling rate to the internal losses, outgoing power can be minimized, resulting in the waves escaping the cavity being out of phase and having the same amplitude as directly reflected waves at the access point. This results in a matched cavity or critical coupling[40,42]. Traditionally, this regime is achieved by driving the cavity with a constant amplitude sinusoidal wave at the resonance frequency of a cavity mode.

Plasma generation hinges on applying a high electric field that triggers substantial gas ionization. As the intensity of this field rises, the time it takes to apply it decreases. In the context of plasma ignition via an electromagnetic cavity, intensifying the resonant electric field can potentially initiate a plasma before reaching a steady state. However, during this phase, the traditionally used critical coupling results in a peak in global reflected power. To address this, we suggest an innovative method that utilizes virtual absorption[6] or, more specifically, virtual critical coupling[7]. Under certain circumstances that we discuss in this work, this approach can eliminate outgoing power during the transient phase, thereby enhancing efficiency.

First, we have designed an aluminum (Al) resonant cavity[11,43], incorporating two metallic concentrators that form a 6 mm gap and align parallel to the electric field, as depicted in Fig. 1a. For optimal control and facilitation of plasma ignition, we have installed a glass tube within our aluminum cavity, strategically placed between the concentrators where the electric field peaks. The tube, with an external diameter of 6 mm and an internal diameter of 5 mm, is sealed at a low pressure of 5.3 mbar (4 torrs) and contains argon. To ensure electromagnetic hermeticity while allowing plasma observation, we have incorporated a Faraday grid into a cavity wall and utilized a camera for monitoring, with its perspective indicated by the violet dashed frame in Fig. 1b.

To maximize electric field concentration between the concentrators, we excite the $TE_{012}$ mode[11] at the real resonance frequency $f_0$. Fig. 1c demonstrates the numerically simulated E-field profile of the $TE_{012}$ mode. We aim to achieve VPA, which is pertinent for optimizing energy stored within the cavity. For this to be effective, we need the mode to be over-coupled, implying that energy leaks at the access faster than dissipating at the cavity walls. This scenario is characterized by an internal decay rate ($\gamma_{int}$, representing dissipative losses) lower than the external decay rate ($\gamma_{ext}$, representing leakage via coupling at the access).

Achieving VPA requires precise determination of three parameters: $f_0$, $\gamma_{int}$, and $\gamma_{ext}$. To this end, we implement the method we have previously established[11], which is rooted in the experimental measurement of the cavity's $S_{11}$ parameter and an analysis of the cavity's response to sinusoidal excitation. Fig. 1d shows the amplitude and phase measurement of this S parameter as a function of frequency. We ensure that the resonance mode $TE_{012}$ is sufficiently isolated from other resonances within the frequency spectrum. A more comprehensive explanation of this method can be found in Methods. From the measurement results, we obtain the following internal and external decay rates: $\gamma_{int}$ = 5.356 $10^6$ $s^{-1}$ and $\gamma_{ext}$ = 90.810 $10^6$ $s^{-1}$. These values correspond to time constants ($\tau = 1/\gamma$) of 186.7 ns and 11.0 ns, respectively. These results indicate that to achieve the VPA regime in our system, the $TE_{012}$ mode must be excited by the exponentially growing signal with complex frequency $f_{zero}$ = 2.416 + j 0.0136 GHz (convention $e^{-j2\pi ft}$). Interestingly, the time constant associated with dissipation losses significantly exceeds that linked to access leakage. Consequently, most of the incident energy is stored during VPA excitation, and only a minimal amount dissipates during excitation. The experimentally obtained pole and zero are presented in Fig. 2a.

Fig. S1 in the Supplementary Materials illustrates the experimental setup. The VPA signal $s_{inc}(t)$ is created by modulating a local oscillator's carrier with an exponentially increasing signal, formulated using an arbitrary waveform generator (AWG). This signal is amplified by a pulsed amplifier with a 6% duty cycle and a 500 μs pulse repetition period to create a sufficiently high field level to ignite plasma inside the tube. An ICCD camera with a short resolution is strategically placed in front of a Faraday grid within the cavity walls (see Fig. 1b) for plasma ignition observation and characterization during excitation. The camera with a 3ns gate time is synchronized with the VPA signal thanks to the AWG trigger with an adjusted delay. Each image is produced from the aggregate of 300 gates. Plasma light intensity is spatially integrated to monitor the discharge's temporal evolution (i.e., the temporal changes of the sum of pixel intensities on the surface depicted in purple in Fig. 1b), termed the "integrated light intensity." For additional details on the experimental setup, please refer to the Methods section and Supplementary Materials.

Figs. 2b,c display the outcomes of the low-power (b) and high-power (c) microwave VPA excitation. Both figures plot the incident and reflected electromagnetic signals, as the oscilloscope captures over time. The results take into account the attenuation and delays encountered by the signals in the measurement channels, using the cavity access as a reference. As such, the cavity receives excitation, which we shape through modulation with suitable exponential growth (at frequency $f_{zero}$ = 2.416 + j 0.0136 GHz), to form the incoming signal $s_{inc\,m}$ (illustrated in dark blue). To achieve the high-power regime in Fig. 2c, the excitation signal is amplified with the pulsed amplifier (see Methods). The final peak voltage of the exponential amplitude at the cavity access attains approximately 3 V and 190 V in the lower-power and high-power regimes, respectively. The corresponding instantaneous powers are 90 mW and 360 W at 50 Ω. In both regimes, the excitation is terminated at time $t_e \approx$ 70 ns.

The measurement results of the reflected signal, $s_{ref\,m}$, without plasma in the resonant cavity, are presented in Figs. 2b,c in yellow. During the excitation period - before time $t_e$ - this signal is a superposition of waves directly reflected at the access and those that leak from the cavity. As anticipated, we observe a substantial absence of reflected waves for most of the excitation period in both low-power and high-power VPA regimes. In the low-power regime (Fig. 2b), any reflection measured is minute and can be considered negligible, demonstrating a clear distinction from the evolving incident signal over time. However, in the high-power regime (Fig. 2c), there is a gradual increase in reflection towards the end of the excitation, between 60 and 70 ns, a phenomenon observed both in the presence and absence of plasma. This suggests that the reflection rise appears ignition-independent and caused by a minor non-linearity of our signal amplifier. Indeed, in the high-power regime, the amplitude of the $s_{incm}$ signal slightly deviates from the ideal one generated by the AWG (shown in light blue).

Following the cut-off point at time $t_e$, the reflected signal, denoted as $s_{ref\,m}$, is solely derived from the cavity's leakage waves through the access, adhering to an exponential decay pattern. This stage signifies the transition of the resonant cavity into a state of free oscillations. This observed exponential decay is distinctly characterized by the S-matrix pole[12,44]. With a time constant of 10.4 ns, the decay predominantly pertains to the pole associated with the $TE_{012}$ mode ($f_{pole}$ = 2.416 - j 0.0153 GHz), given that resonances remain relatively isolated within the operational frequency band.

The emergence of plasma ignition hinges on the initial presence of electrons or ions dwelling within the cavity due to cosmic rays[45] or stemming from the effect of previous pulses in a pulsed excitation scenario. This latter phenomenon, known as the "memory effect"[46,47], will be further explored in what follows. These initially present charged entities can be set into motion under the influence of the microwave electric field, causing swift ionization or an electron avalanche. However, shielding metallic walls in our resonator can impede the formation of initially charged particles. To remedy this, our experiments employed a triggering light source that facilitated forming initial ions solely for the first pulse (see Methods). In this manner, we could effectively regulate the initial conditions and eliminate the variables that could otherwise cloud our results.

The results shown in Fig. 3a were obtained using a final peak voltage of 190 V (measured at cavity access) with a repetition period of 500 μs (interval between two successive excitations). The signal observed reminds the one in Fig. 2c (without plasma). Yet, in this case, we have measured the reflected signal produced when plasma is generated during the VPA excitation (in green). We have also plotted the reflected signal when plasma is not present (in yellow) for comparison. Except for the ten nanoseconds before $t_e$ (discussed above), and just like the situation with no plasma, we note that there is no reflected wave until the completion of the excitation, which is a signature feature of VPA.

Apart from ionization, the interactions between particles also induce the excitation of atoms or molecules. Consequently, the subsequent de-excitation causes light emission from the plasma, which is then captured by our camera. This process serves two purposes: it triggers the ignition of the plasma and visually confirms its presence and intensity. We employ an intensified charge-coupled device (ICCD) camera renowned for its superior temporal resolution to monitor plasma initiation during excitation. The camera allows tracking of the time-dependent evolution of the accumulated light intensity, which we have presented in Fig. 3a by the red dashed curve with circular markers. In the inset, we show the image captured by the camera at peak light intensity. The existence of plasma in the tube between the concentrators is evident in this image. The pattern is attributable to the intervening grid between the plasma and the camera. Background noise was successfully eliminated from the image by conducting an initial measurement without plasma. We then reprocessed the intensity level in violet - a hue characteristic of argon plasma and perceptible to the naked eye. The tube's boundary is delineated with white dotted lines. We note a rapid surge in light intensity - approximately 15 ns - just before the cessation of excitation, which coincides with a highly intense electric field. It is worth mentioning that the light intensity increased by a factor of 4000 from the initiation of excitation to the peak intensity. As the energy sustaining the discharge diminishes, we observe a corresponding gradual decrease in light intensity. Thus, the experiment demonstrates the absence of reflection and plasma ignition during excitation, suggesting the potential for utilizing VPA to achieve reflectionless plasma ignition.

As previously discussed, one key phenomenon to take into account in the context of pulsed discharges is the 'memory effect.' This concept elucidates the influence of one pulse on the subsequent one, as evidenced by various studies[46,47]. The memory effect is intrinsically tied to the enduring presence of entities such as charged particles and excited species from one discharge to the next. For instance, in the results shown in Fig. 3a, which reflect a 500 μs repetition period, the existence and generation of electrons by microwaves do not notably impact the electromagnetic conduct of the cavity during the excitation phase. This leads to a reflectionless plasma ignition, granted that the electron density remains adequately low before and during the excitation. However, when the repetition period shortens, the electron density might increase to a point where it significantly affects the electromagnetic performance of the cavity.

To illustrate this, Fig. 3b presents results from a 200 μs repetition pulse sequence. Here, the previously achievable reflectionless excitation via VPA becomes untenable. The observed leakage signals before the excitation cut-off exhibit a pronounced difference in amplitude when compared to the presence versus absence of plasma (Fig. 3a). Plasma presence results in part of the incident energy, which accumulates within the cavity during excitation not being discharged afterward, a contrast to the situation when plasma is absent. This suggests that plasma absorbs some of this energy, positioning it as a lossy medium and inducing a fluctuation in the cavity's internal dissipation factor, $\gamma_{int}$. Additionally, the plasma density may dynamically evolve during the excitation process. Hence, in the case of this type of transient plasma discharge, the medium experiences a change in its properties over time, specifically in its electromagnetic characteristics. The associated S-matrix zero is affected, potentially evolving in time alongside the plasma. The energy dissipation from the cavity post-excitation is relevant in this context. The signal's amplitude deviates from the typical exponential decay pattern, a phenomenon expected when the propagation medium is stationary. Like the S-matrix zero of the mode under consideration, the cavity's poles shift with the presence of plasma. This dynamic behavior of the medium introduces challenges and necessitates the development of a reflectionless excitation method. This method, inspired by VPA, should consider this non-linearity. Essentially, it implies defining a frequency, $f_{zero}(t)$, that evolves synchronously with the plasma discharge. This approach appears logical, especially considering the unchanging interference principle governing the cavity access reflection during excitation.

**Conclusion**

In light of the findings of this work, it has become evident that plasma generation, crucial for a myriad of industrial applications, can be significantly optimized using the innovative concept of VPA. This novel approach mitigates the longstanding challenges in plasma ignition, such as energy reflections and component damage, by facilitating the reflectionless excitation of a resonator, thereby ensuring the complete storage of the incident energy during the excitation process. Our methodology departs from conventional techniques by meticulously designing the temporal profile of the incident wave, minimizing energy reflections during ignition. The empirical results presented herein underscore VPA's capability to successfully ignite a plasma discharge within a gas-filled resonator, signaling VPA's potential to greatly enhance the durability and efficiency of plasma ignition systems. Furthermore, the data suggest that VPA can be leveraged to achieve reflectionless plasma ignition during excitation with minimal dissipation, dedicating almost the entire incident energy to ignition. Reflection elimination is realized not through external electronic or mechanical circuits but by strategic control over the incident waveform. These findings open the door to developing methods for plasma ignition without reflection during both ignition and plasma sustenance periods. On a broader scale, this research affirms the wide-ranging possibilities presented by waveform control within a plasma context, a control mechanism recently proposed for spatiotemporally steering plasmas[48,49]. The results of this study serve as a stepping-stone for substantial breakthroughs in plasma technology, offering a sturdy solution to enduring challenges in plasma ignition and underlining the untapped potential of waveform control.

**Methods**

**Experimental S-matrix zero extraction.** Determining the complex excitation frequency associated with VPA, also known as the S-matrix zero[12,50], can be challenging in experimental resonant cavities. Recently, we proposed a method based on analyzing the temporal response of the cavity to harmonic excitation, enabling us to extract this S-matrix zero[11]. It is noteworthy that this method applies explicitly to over-coupled modes and only when they are sufficiently distanced from other resonances, which is the case in this study. In our single-port cavity experiment, we employ a Vector Network Analyzer (VNA) to measure the $S_{11}$ across the relevant frequency band. This methodology permits the numerical construction of the cavity's response to an arbitrary incident signal, achieved through a convolution product. Subsequently, we analyze the frequency of the pole (the leakage signal after the excitation cut-off that delineates the resonator's behavior during free oscillations) to determine the real resonance frequency. By examining the transient and permanent time regimes of the reflected signal for a harmonic excitation at this real frequency, we measure two key quantities - the instant of reflection cancellation during the transient period[7] and the value of the reflection coefficient at the steady state. Inputting these two quantities into a system of two equations with two unknowns, derived from the temporal coupled-mode theory[40], enables us to derive the internal decay rate $\gamma_{int}$ and external decay rate $\gamma_{ext}$ of the cavity for the excited resonance. At this juncture, we possess the three parameters that characterize the complex frequency: $f_{zero} = f_0 + j\,(\gamma_{ext} - \gamma_{int})/2\pi$ ($e^{-j2\pi ft}$ convention).

**Experimental high-power setup.** Our research relies heavily on the manipulation and measurement of signals. An incoming signal is initially molded by modulating a carrier generated by the Tektronix TSG4104A Local Oscillator (LO) generator and an exponential envelope produced by the Keysight 33600A Arbitrary Waveform Generator (AWG). This modulation occurs via the 'I' channel of the Polyphase I/Q modulator QM2040A. The AWG's maximum output amplitude is adjusted for the highest possible incident power. To provide the cavity with a power-packed signal that can spark plasma, we use the Travelling Wave Tube Amplifier (1.9 kW pulsed 2-8 GHz PTC7353) by TMD Technologies. We have also installed a pre-amplification channel upstream to bolster the amplifier's input power. Another AWG, the Rigol DG400, synchronizes the signal emission and amplification in pulsed mode. This AWG is instrumental in adjusting the repetition period of the excitation. Additionally, we use a circulator in conjunction with a 300 W stationary matched load to safeguard our amplification system from wave return from the cavity. We employ a directional coupler to capture attenuated signals that are incident on and reflected from the cavity for measurements. These measurements are conducted using a Keysight MSO9254A oscilloscope. As plasma measurement necessitates repeatability, we measure, and record averaged electromagnetic signals on the oscilloscope, thus reducing measurement noise. We characterize the measurement channels of the incident and reflected signals based on attenuation and time delay to correlate our results with events at cavity access.

We use the Princeton Instruments PI-MAX 4: 1024f high temporal resolution ICCD camera for the photographic measurements. During the excitation phase, the camera is triggered to collect light data from the plasma for 300 repetitive discharges over 3 nanoseconds at the same time delay before transitioning to the subsequent 3 nanoseconds delay. The Rigol AWG simultaneously triggers the emission of the excitation by the Keysight AWG and the amplification. Furthermore, this second AWG also synchronizes the emission of the incident signal and camera trigger. We can harmonize our measurements in time by measuring this synchronization signal alongside the camera's monitoring signal (when it initiates recording) and the incident and reflected signals.

The first ignition process is eased by illuminating the cavity's interior via the grid with a low-energy visible light source, such as a white LED emitting a spectrum between 400 and 700 nm and a variable luminous flux ranging from 20 to 200 lumens. Given the cavity's metallic and entirely opaque nature, this auxiliary light source guarantees an adequate electron presence within the discharge region, thereby minimizing statistical fluctuations in the initial breakdown field, as indicated in references[51–54]. For subsequent breakdowns, the light source becomes unnecessary due to the residual charges (or metastable argon particles) from the prior breakdown, which persist at the end of the previous pulse, showcasing a memory effect.

## Acknowledgments

We would like to thank CEA Gramat for the loan of the pulsed amplifier.

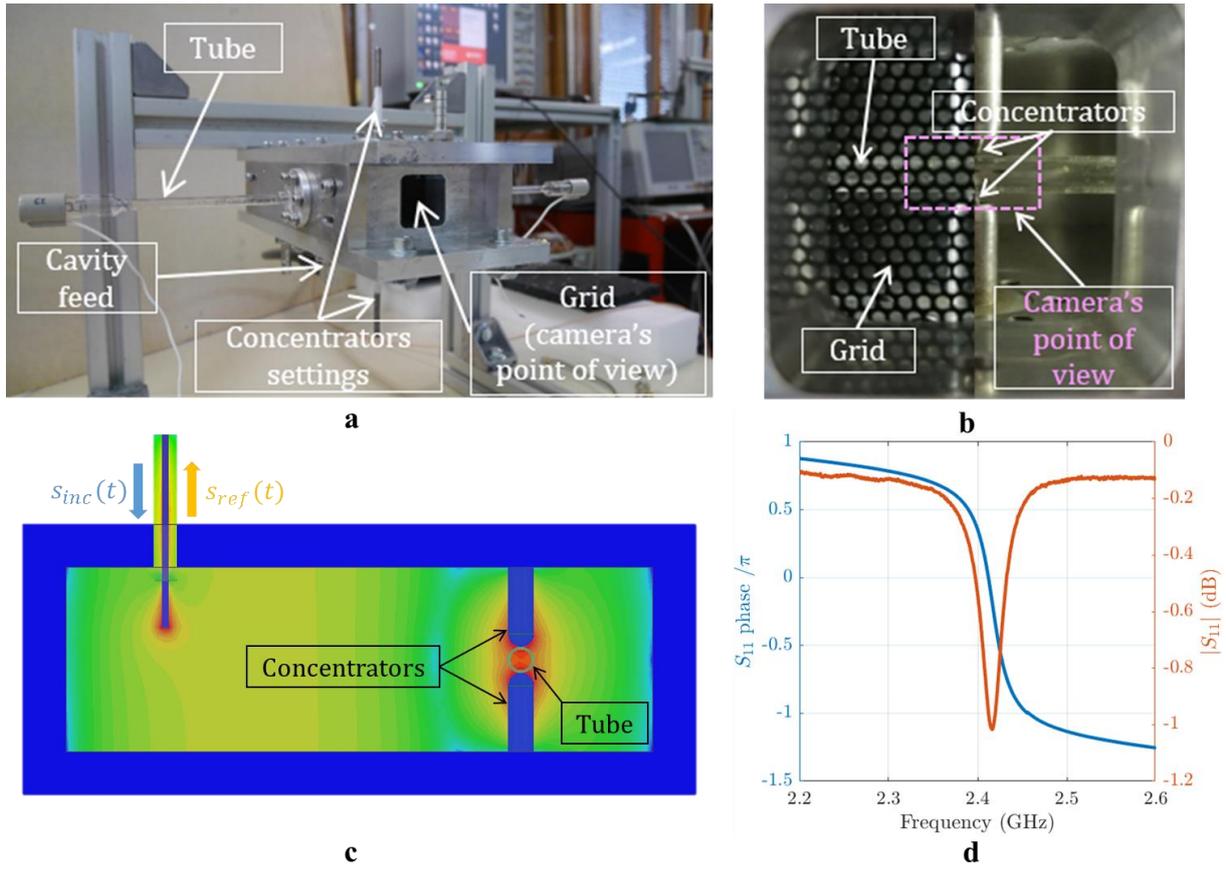

*Figure 1. a,b, Photographs of the cavity. External (a) and internal (b) views (with and without the Faraday grid). c, Numerically simulated E-field profile of the $TE_{012}$ mode. d, $S_{11}$ as a function of frequency, measured at the vector network analyzer.*

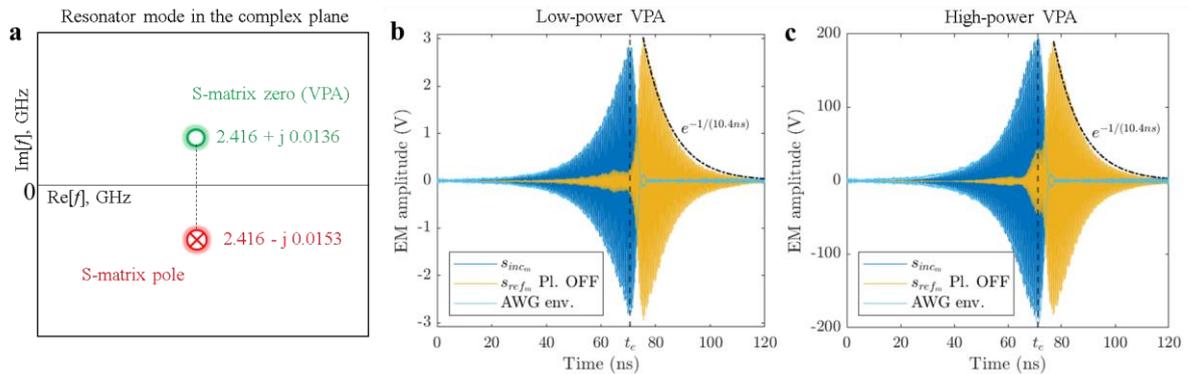

*Figure 2. a, Experimentally deduced position of poles and zeros of the resonant mode under consideration. b,c, Low-power (b) and high-power (c) VPA without plasma: measurements of the incident and reflected microwave signal amplitudes with the oscilloscope (no plasma, 500 μs repetition period). These results are averaged over 64 measurements. The raw signal captured on the oscilloscope for a single pulse is provided in the Supplementary Material.*

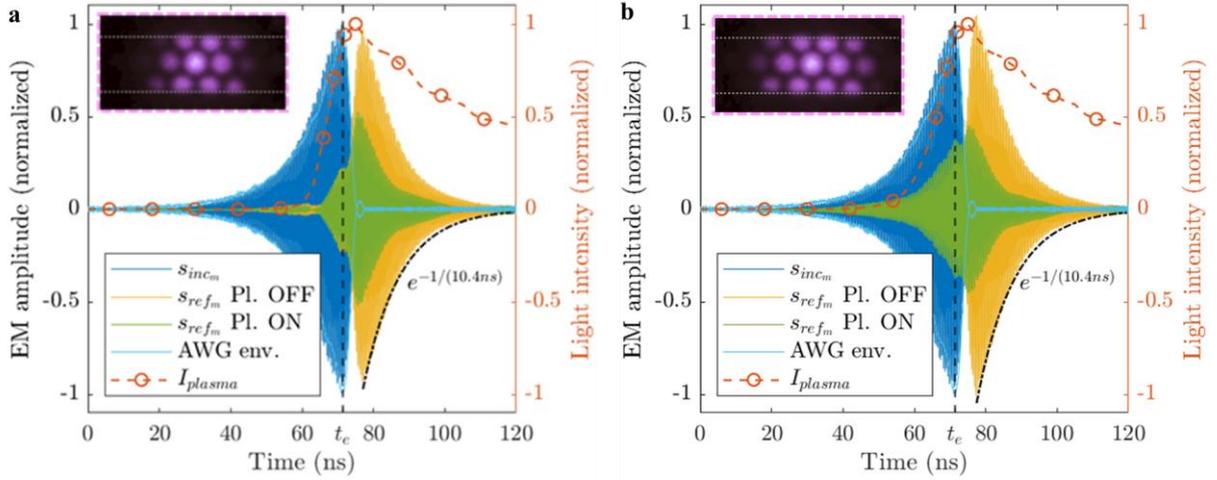

*Figure 3.* *Plasma ignition by high-power VPA.* ***a****, Measurements of the incident and reflected microwave signal amplitudes (plasma OFF and ON, 500 μs repetition period). The integrated light intensity of the plasma is plotted over time. Inset: the image of the plasma captured by the camera for an integration gate of 3 ns around the light intensity peak.* ***b****, Alteration of S-matrix zero by plasma: measurements of the incident and reflected microwave signal amplitudes with the oscilloscope (plasma OFF and ON - 200 μs repetition period). These results are averaged over 64 measurements. The light intensity is normalized to the maximum of the previous case (500 μs repetition period) and has a peak of 5 % higher. Note the wider plasma spread in the tube compared to this first case in the inset.*